# Fabrication of Fibers with Complex Features Using Thermal Drawing of 3D-Printed Preforms

Ali Anil Demircali [a,†], Jinshi Zhao [a,b,†], Ayhan Aktas [c], Mohamed EMK Abdelaziz [b], and Burak Temelkuran [a,b,*]

[a] Department of Metabolism, Digestion, and Reproduction, Faculty of Medicine, Imperial College London, London W12 0NN, UK.

[b] The Hamlyn Center for Robotic Surgery, Imperial College London, London SW7 2AZ, UK.

[c] Mechatronics in Medicine Laboratory, Hamlyn Center, Imperial College London, London SW7 2AZ, UK

[*] Corresponding Author: b.temelkuran@imperial.ac.uk

[†] These authors contributed equally to this work.

**Abstract:** High-aspect-ratio polymer materials are widely utilized in applications ranging from everyday materials such as clothing to specialized equipment in industrial and medical fields. Traditional fabrication methods, such as extrusion and molding, face challenges in integrating diverse materials and achieving complex geometries. Additionally, these methods are limited in their ability to provide low-cost and rapid prototyping, which are critical for research and development processes. In this work, we investigated the use of commercially available 3D printers to fabricate fiber preforms, which were subsequently thermally drawn into fibers. By optimizing 3D printing parameters, we achieved the fabrication of fibers with diameters as small as 200 µm having complex shapes, with features down to a few microns. We demonstrated the versatility of this method by fabricating fibers from diverse set of materials, such as fibers with different stiffnesses and fibers with magnetic characteristics, which are beneficial for developing tendon-driven and magnetically actuated robotic fibers. In addition, by designing novel preform geometries, we produced tapered fibers and fibers with interlocking mechanisms, also tailored for use in medical steerable catheter applications. These advancements highlight the scalability and versatility of this approach, offering a robust platform for producing high-precision polymer fibers for diverse applications.

# 1. Introduction

High-aspect-ratio polymer materials are integral to modern technologies due to their versatility, enabling flexibility with intricate geometries. They are used in applications ranging from everyday textiles to cutting-edge fields such as optical transmission [1,2], and medical instruments such as catheters [3,4]. Realizing these advanced applications requires the development of thin and long materials that meet demanding criteria for structural complexity, material functionality, and scalable production.

To meet these requirements, polymeric material fabrication methods, such as extrusion and molding, have been developed to address specific needs [5]. Among these methods, extrusion is an effective technique for scaling production while being compatible with handling a range of designs [6,7]. However, this process is fundamentally constrained by the need for specialized dies that are both time-intensive and costly to produce [8–12]. Additionally, the fabrication of multi-material products encounters further challenges, such as viscous encapsulation and material incompatibility due to differences in their thermomechanical properties [13,14]. Another alternative method, molding, excels in the production of intricate geometries and achieves better precision on a miniature scale [15]. Despite these advantages, molding has its limitations, including high initial tooling costs, prolonged lead times, and inefficiencies in fabricating continuous structures with long lengths [16,17]. While extrusion and molding address specific manufacturing needs, their limitations underscore the demand for a scalable method capable of producing thin, long, and flexible structures with multiple materials and precise geometries.

Thermal drawing provides a solution for fabricating high-aspect-ratio polymer materials (known as fibers) with intricate cross-sections while maintaining their structural integrity. Thermal drawing is a fabrication process that transforms macro-structured preforms into fibers while preserving their cross-sectional features [18–20]. The preform is typically a macroscopic structure measuring a few centimeters in diameter and tens of centimeters in length. The preform, which can be made of amorphous polymers, is placed in a draw tower and heated to its viscoelastic state. Axial tension is then applied to produce thin, continuous fibers that can extend up to hundreds of meters in a single draw [21–23]. Preform fabrication is one of the most crucial steps in fiber production, as it significantly influences the fiber's quality, overall fabrication time, material usage, and both financial and labor costs. Particularly for research and product prototyping purposes, using preform fabrication methods that allow for rapid iteration can provide timely product updates and experiential optimization.

The fabrication of polymer preforms can be achieved through various methods, each offering advantages and challenges. Optical fibers have a relatively simple structure and are fabricated from materials with a cylindrical geometry. However, with the increasing demands of functional fibers, new preform fabrication methods have been developed to achieve fibers with more complex cross sections. The stack-and-draw method, assembles rods, tubes, or plates into a desired preform structure [24,25]. Additionally, thin-film rolling offers a unique approach by rolling polymer films, consolidating them under vacuum, and enabling the creation of hollow-core designs and nano-layered structures, such as photonic bandgap fibers [26–28]. However, this method is suitable for cylindrical preform shapes, and it is less suitable for irregularly shaped designs. Molding is another technique that involves melting pellets within molds to produce intricate shapes. This process ensures precise geometries by replicating the mold's structure during solidification [29–31]. Despite its precision, molding faces challenges such as high tooling costs and difficult disassembly of molds with complex shapes [32–34]. As another alternative, bulk material machining, such as drilling polymer rods, is another method primarily used for producing multi-lumen tubing. The preforms fabricated using this method are limited to simpler designs and require high-precision machining mechanisms, such as using computer numerical control (CNC) machines [35,36].

Among these methods, 3D printing-based preform manufacturing stands out by its ability to create complex, multi-material geometries without the need for custom molds or extensive manual handling [37–40]. This technique facilitates the incorporation of intricate features, such as hollow cores, small channels, and arbitrary cross-sectional shapes, which are difficult or impossible to achieve using traditional methods. While early attempts to use 3D printing in fiber manufacturing focused primarily on optical fibers [41–44], recent demonstrations have shown promise in expanding their use to medical field [38,45,46]. While these studies focused more on specific applications, there is a need for a systematic approach to improving the process by investigating material compatibility, structural integrity, and various manufacturing parameters [37,47,48].

Here, we present a hybrid approach that integrates 3D printing and thermal drawing to address the limitations of existing fabrication techniques, particularly in achieving complex geometries and integrating diverse materials (**Fig. 1**). By systematically investigating 3D printing parameters, and thermal drawing conditions, we demonstrated the fabrication of advanced fiber designs. First, we fabricated fibers with intricate cross-sectional geometries, including rabbit, butterfly, and star-shaped designs. The fiber diameter is down to 200 µm, demonstrating the scale-down capability of preserving cross-sectional features. Second, we thermally drew semi-crystalline polypropylene (PP) fibers, marking the first demonstration of this material in thermal drawing process. PP's combination of low stiffness, toughness, chemical resistance, and superior flexibility compared to polycarbonate (PC), making it ideal for applications like medical catheters. Third, we developed preforms with longitudinal periodic thickness variations, which were subsequently drawn into tapered fibers. Additionally, we fabricated magnetic fibers from fully 3D-printed preforms, enabling magnetic actuation for robotic applications. These advancements highlight the versatility of combining 3D printing and thermal drawing to fabricate fibers, addressing longstanding challenges in polymer fiber fabrication while paving the way for new applications across healthcare, robotics, and sensing technologies.

## 2. Materials and Methods
### 2.1 Preform Fabrication and 3D Printing Parameters
Preform designs were created using SolidWorks (Dassault Systems, France) CAD software and fabricated with a Fused Deposition Modeling (FDM) printer (Ultimaker 3+ Extended, Ultimaker BV, Netherlands). The printer employed a 0.4AA print core for non-abrasive plastics. Fabrication parameters included 100% infill density, triangular infill pattern, and a layer thickness of 0.08 mm. Extrusion speeds were adjusted based on material: 70 mm/s for PC, 35 mm/s for PP, and 50 mm/s for mPLA. Preforms, measuring 25 mm in diameter and 150 mm in length, were fabricated from polycarbonate (Ultimaker PC Transparent, Netherlands), polypropylene (PP 3D FilaPrint, England), and magnetically loaded polylactic acid (Proto-Pasta Iron Composite, USA). To ensure adhesion, an adhesive layer (DIMAFIX, 3D GBIRE, United Kingdom) was applied to the glass build plate. Fabrication conditions were modified based on material properties: for PC and mPLA, an upper enclosure and door (Accante cover, 3D GBIRE) maintained chamber temperature, whereas for PP, the enclosure and door were removed for faster cooling. Nozzle ($T_n$) and bed ($T_s$) temperatures for each material are detailed in Table 1.

Table 1 Temperature setting for printing.

|      | $T_n$ (°C) | $T_s$ (°C) |
|------|------------|------------|
| PC   | 270        | 107        |
| PP   | 230        | 85         |
| mPLA | 210        | 85         |

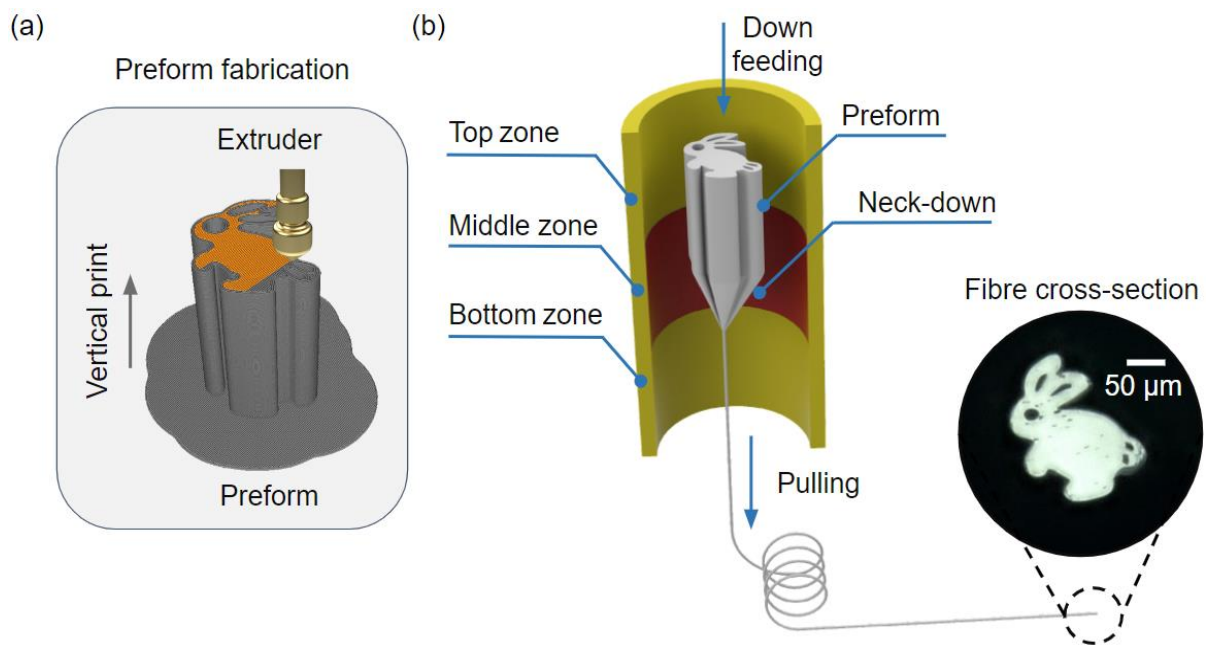

**Fig. 1 Schematics of the preform fabrication and thermal drawing process**. (a) A fiber preform is vertically printed using Fused Deposition Modeling (FDM) 3D printing technology. (b) The preform is fed into the three-zone furnace of the draw tower, where the fiber diameter and tension are monitored in real-time using laser displacement and force sensors. Subsequently, the preform is drawn into a fiber while maintaining its cross-sectional shape.

## 2.2 Thermal Drawing
The 3D-printed preform was secured to a preform holder made from 30% glass fiber-reinforced PEEK (Ketron PEEK Glass Fiber Reinforced Rod, 20 mm diameter, Bay Plastics Ltd, UK), chosen for its high melting point of 340°C. This holder was mounted on the motorized down-feed stage of a draw tower to control the descent of the preform into a three-zone radiative cylindrical furnace. Each furnace zone, measuring 100 mm in height, was independently temperature controlled. The top zone preheated the preform to near its glass transition temperature (Tg), the middle zone provided the maximum temperature to achieve drawable viscosity, and the bottom zone stabilized the fiber dimensions and cross-sectional features by cooling it to ambient temperature.

At the start of the process, approximately 50 mm of the preform's bottom section was positioned in the middle heating zone. Weights (specified in Table 2) were attached to the preform's lower end to generate a consistent downward pulling force. This force facilitated the neck-down process, where the preform's diameter gradually decreased as it elongated into a fiber. Heating softened the material, allowing gravity and applied tension to elongate the preform into a continuous fiber. Furnace temperatures for each material were set according to Table 2, with heating durations ranging from 30 to 60 minutes, depending on the material's thermal properties.

Fiber dimensions were monitored in real-time during the process using a 2-axis laser displacement sensor. The fiber diameter was controlled by maintaining a preform feed rate of 2 mm/min and adjusting the winding speed to achieve the desired dimensions. The final fiber diameter was determined using the equation:

$$r_d = r_p \sqrt{\frac{v_d}{v_w}} \quad (1)$$

where $d_f$ is the fiber diameter, $d_p$ is the preform diameter, $v_d$ is the preform feed rate, and $v_w$ is the winding speed. Tension on the fiber was continuously measured and maintained below 100 g to ensure dimensional consistency and avoid fiber breakage.

Table 2 Drawing Temperatures

|  | $T_t$ (°C) | $T_m$ (°C) | $T_b$ (°C) | $m_h$ (gr) |
| --- | --- | --- | --- | --- |
| PC | 140 | 200 | 85 | 20 |
| PP | 100 | 135 | 65 | 10 |
| PC+mPLA | 140 | 210 | 95 | 25 |

## 2.3 Fabrication and characterization of PC and PP fibers

The bending behavior and mechanical properties of PC and PP fibers were evaluated through tendon-driven force characterization. Fibers with identical cross-sectional designs were prepared in lengths of 6 cm, 8 cm, and 10 cm. A 100-µm molybdenum wire was threaded through the side lumens of each fiber, with the distal end of the wire secured and the proximal end exposed for connection to a force sensor. The force sensor was mounted on a manual linear stage, allowing precise control and measurement of the pulling force applied to the wire.

To ensure uniform initial conditions, each fiber was positioned in a fixed alignment jig with predefined anchor points. The proximal end of the fiber was clamped securely to prevent movement, while the distal end was aligned horizontally. Fibers were placed on a flat, marked reference surface to maintain consistent orientation within the field of view of a digital microscope (VHX-6000, Keyence, Japan). Bending displacements were recorded using the microscope equipped with a 5x objective lens. Images of the fiber curvature were analyzed using a MATLAB program that digitized and reconstructed the fiber shapes from ten evenly spaced points along the curvature. The relationship between the applied pulling force and the resulting bending angle was computed to quantify the mechanical response of PC and PP fibers under applied forces.

### 2.4 Iron-filled Fiber Characterization

The iron-filled PLA composite is ferromagnetic, and the magnetic response of the fiber made from ferromagnetic PLA (mPLA) was characterized by measuring the displacement of the fiber tip in response to a permanent magnet (Neodymium N35 grade block magnet, 40 mm x 25 mm x 30 mm). The magnet was securely mounted to a manual stage (LT3/M, Thorlabs, USA) for precise positioning relative to the fiber. A digital camera (VHX-6000, Keyence, Japan) equipped with a 10x objective lens was aligned orthogonally to the experimental setup to monitor and record the fiber's motion.

To systematically evaluate the magnetic response, the distance between the magnet and the fiber tip was varied in 1 mm increments, beginning with direct contact and extending to a maximum separation of 20 mm. At each position, the displacement of the fiber tip was recorded using the digital camera. The recorded displacement data were analyzed to quantify the fiber's magnetic response as a function of the magnet's proximity.

## 3. Results and Discussion
### 3.1 Fiber Characterization
#### 3.1.1 Effect of Printing Layer Thickness

To investigate the effect of printing layer thickness on fiber diameter consistency, we explored a range of layer thicknesses from 50 µm to 250 µm, in 50 µm increments. Hollow-core PC preforms with an outer diameter (OD) of 20 mm, an inner diameter (ID) of 6 mm, and a length of 100 mm were used in the experiments. During the draw, a consistent fiber diameter of 1 mm was targeted, with constant feed (2 mm/min) and draw speeds maintained across three tests.

The relationship between printing layer thickness, printing time, and fiber diameter fluctuation is illustrated in Fig. 2. Reducing the printing layer thickness significantly decreased fluctuations, improving fiber consistency, while increasing the layer thickness resulted in greater fluctuations, following an exponential trend (**Fig. 2a**). For instance, a 50 µm layer thickness achieved a standard deviation of ±19 µm, whereas increasing the layer thickness to 250 µm caused the standard deviation to rise exponentially to ±400 µm, corresponding to a fluctuation of 2% to 44% around the targeted fiber diameter (**Fig. 2b**). Additionally, printing time was inversely related to layer thickness. For example, increasing the layer thickness from 50 µm to 100 µm approximately halved the printing time, reducing it from 656 minutes (about 11 hours) to 337 minutes (about 5.5 hours) (**Fig. 2c**).

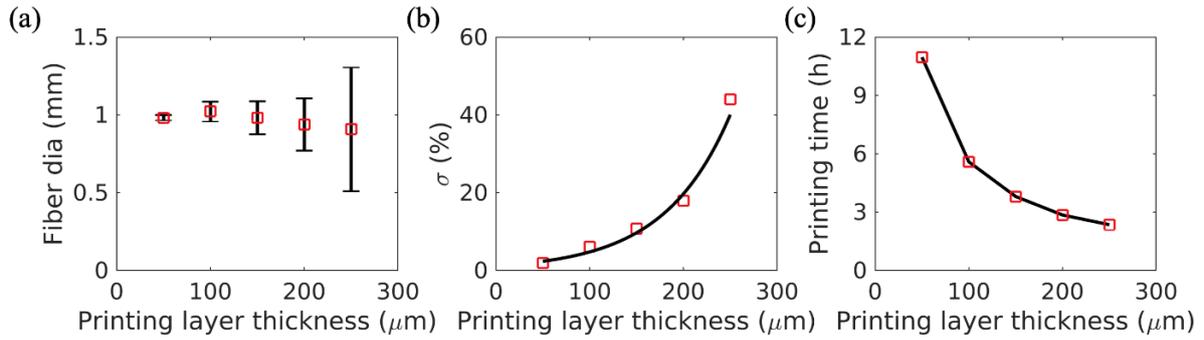

**Fig. 2 Evaluation of fiber diameter consistency and printing layer thickness effects.** (a) Variation in fiber diameter with different printing layer thicknesses, targeting 1 mm diameter, showing increased fluctuation at higher printing layer thicknesses. (b) Standard deviation in fiber diameter as a function of printing layer thickness, illustrating an exponential increase in fluctuation. (c) Printing time is a function of printing layer thickness, highlighting the trade-off between precision and time.

The results highlight the exponential growth in standard deviation with increasing layer thickness, emphasizing the trade-off between precision and efficiency in the manufacturing process. While a lower layer thickness is beneficial for fiber dimension accuracy, it results in longer printing times due to the need for more printing layers. Based on this, we chose an 80 µm layer thickness for our 3D printed preform fabrication to ensure that the standard deviation of fiber dimensions remained within ±50 µm. This choice provides a practical balance between dimensional accuracy and reasonable printing duration (approximately 6-7 hr per preform), facilitating the production of high-quality fibers.

### 3.1.3 Fabrication of Various Fiber Designs

We 3D printed 150 mm long PC preforms with cross-sectional geometries shaped as a rabbit, butterfly, and star (**Fig. 3a,** top view **Fig 3c**). These preforms were thermally drawn into fibers while maintaining recognizable shapes (**Fig. 3d**). For the rabbit-shaped fibers, the vertical cross-sectional length (ear to foot) was drawn to diameters ranging from 2.53 mm to 0.19 mm, corresponding to draw ratios between 1:12.6 and 1:168 (**Fig. 3b**). After collecting the microscope images, we manually overlapped them with their original designed shapes and measured the maximum deviation between them. The rabbit fibers exhibited minimal distortion, with the largest deviation of 25 µm occurring at the ears (**Fig. 3e**). Butterfly-shaped fibers preserved their features across draw ratios up to 1:240 and diameters as small as 0.12 mm, with the largest deviation of 10 µm observed at the wing tips. The star-shaped fibers retained five equidistant holes (50 µm each) surrounding a central hollow star at draw ratios up to 1:120 and diameters of 0.30 mm.

The cross-sectional views of the fibers (**Fig. 3e**) demonstrate the precision of the thermal drawing process, with red outlines indicating the uniformly scaled down preform shapes. For the rabbit-shaped fibers, the body region retained its shape well, while the ears and tail experienced slight deformation. In the butterfly-shaped fibers, the thin antennae regions exhibited minor distortion, and the broader wings showed slight curvature at the tips. In the star-shaped fibers, the five equidistant holes, although slightly distorted, maintained consistent spacing and overall shape.

COMSOL simulations (**Fig. 3f**) displayed the stress distribution patterns observed in the fibers during thermal drawing process. From the simulation results, high stress was concentrated at thin or protruding features, such as the rabbit's ears and tail, the butterfly's antennae, and the edges of the star. The rabbit body and butterfly wings exhibited lower stress due to broader geometry, resulting in better preservation of these regions during drawing. These results present how stress distributions and initial preform quality govern the final fiber geometry. The thermal drawing process scaled intricate shapes effectively, but imperfections at the preform level and stress-prone regions limited dimensional accuracy in some areas. By optimizing preform design and reducing initial defects, these limitations can be mitigated to achieve improved precision for complex cross-sectional fibers.

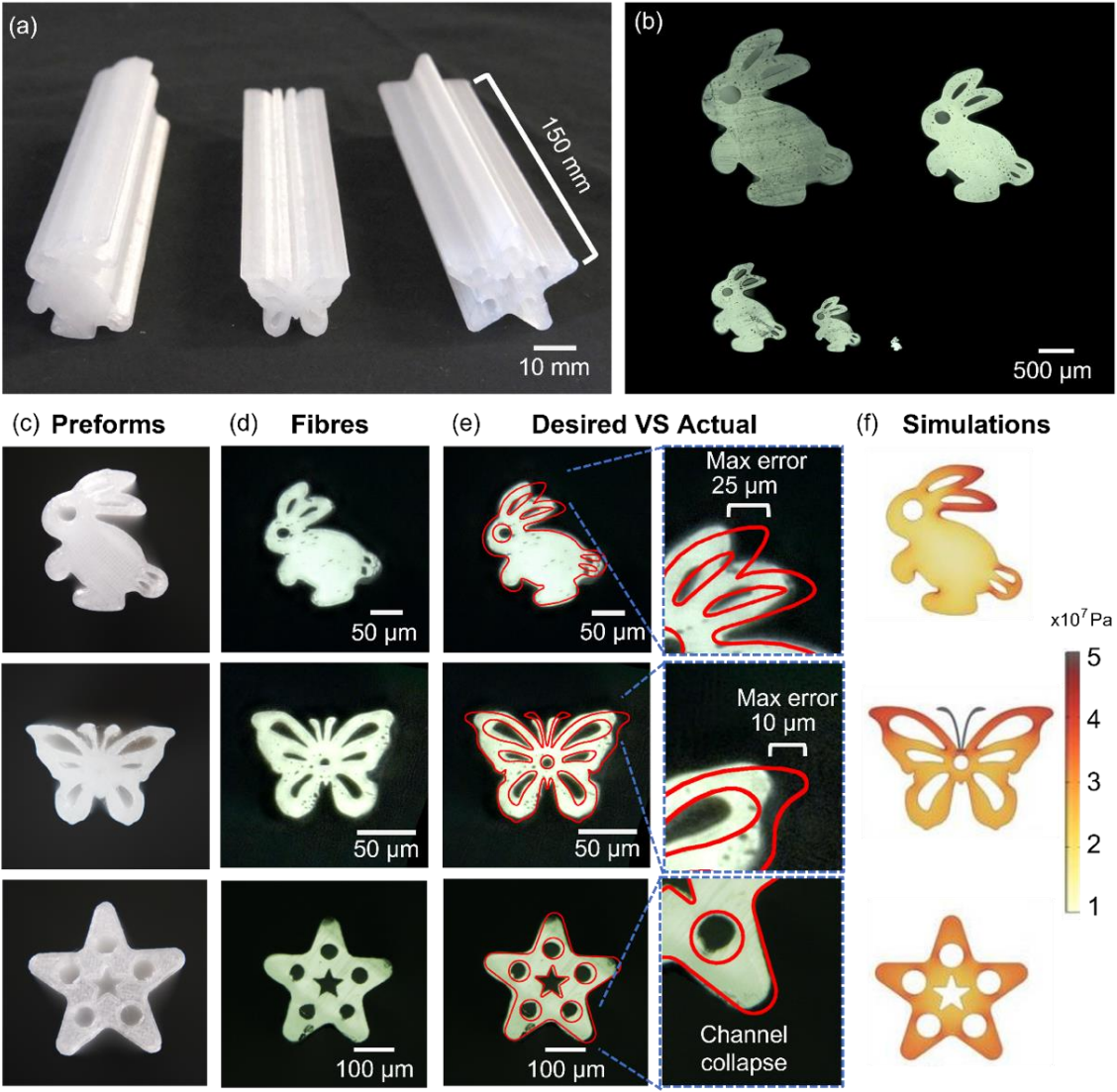

**Fig. 3 Preform structures and their corresponding fibers.** (a) Isometric view of 3D-printed preforms (150 mm in length) with rabbit, butterfly, and star-shaped cross-sections. (b) Top view showing the scale-down effect for rabbit-shaped preforms, demonstrating scalability while preserving intricate cross-sections. (c) Top view of preform cross-sections for rabbit, butterfly, and star shapes. (d) Top view of fibers drawn from these preforms, illustrating the preservation of cross-sectional detail at a scale of 50 µm. (e) Comparison of desired versus actual fiber shapes, with red outlines indicating the target cross-sections. The maximum errors for each shape are provided, with the star shape showing channel collapse. (f) Simulation of stress distribution across the cross-sections of fiber with each shape, with red regions indicating the highest stress and yellow regions indicating the lowest stress.

## 3.2 Advanced Functional Fibers

We fabricated various functional devices by combining 3D printing with thermal drawing technology, presenting the versatility and practicality of this approach through: a) designing preforms with bespoke geometries in both cross-sectional and longitudinal directions to fabricate fibers for medical catheter applications (Sections **3.2.1** and **3.2.2**); b) creating preforms from different materials to fabricate fibers with varying stiffness and magnetic characteristics for robotic actuation purposes (Sections **3.2.3** and **3.2.4**); c) integrating 3D printing with other preform fabrication techniques to explore potential optical fiber fabrication (Section **3.2.5**).

### 3.2.1 Jigsaw Puzzle Piece Shape Design

To demonstrate that this approach can manufacture functional devices with complex geometries, we explored a research focusing on the complex geometric design of high-aspect-ratio materials [46]. The Programmable Bevel-Tip Needles (PBNs) are designed for enhanced precision in neurosurgery, which design incorporates multiple bevel-tip segments with interlocking structures to hold them together. This design allows dynamic offset adjustment between the segments and using tissue interaction forces the catheter's tip to bend in a specific direction when segments are advanced (**Fig. 4a**). In the past, the extrusion process employed for manufacturing PBN catheters achieved the smallest diameter in 2.5 mm [46,49]. In contrast, by employing 3D printing and thermal drawing techniques, the diameter of the catheter has been reduced to 1.3 mm [46], enhancing their suitability for intricate neurosurgical applications.

We used optimized printing and drawing parameters to fabricate PBN with four segments (**Fig. 4b–e**). The 3D-printed preform (**Fig. 4b**) and its corresponding fiber (**Fig. 4c**) demonstrate the quality of shape preservation in a complex cross-sectional design. Figure 4d presents the cross-section, showing four interlocked segments, while Figure 4e illustrates that these segments can smoothly slide relative to each other. Our approach enables the efficient and rapid fabrication of fibers with complex geometries, potentially advancing surgical applications that require precision and multifunctionality.

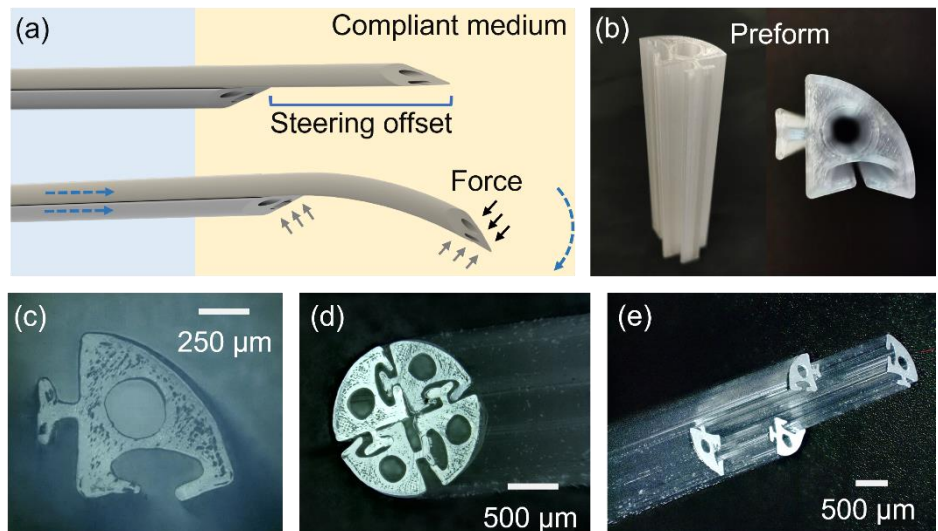

**Fig. 4 Programmable bevel Tip Needles.** (a) Steering concept for 4-Segment PBN (b) 3D printed PBN fiber preform. (c) Cross section view of one segment of PBN with working channel 0.21 mm. (b) Cross Section view of 4-Segment PBN with diameter 1.3 mm. (d). Four segments smoothly slide relative to each other.

### 3.2.2 Preforms with longitudinal periodicity for tapered fibers

The fabrication of a catheter that combines a rigid shaft with a partially steerable tip is typically achieved through either softening the catheter tip by post machining or reinforcing the rest of the catheter with a more rigid structure to enhance stiffness while maintaining flexibility at the tip [50]. To demonstrate the efficiency of our approach, we fabricated fibers with variable diameters along their length to create a tendon-driven catheter platform. This design enables a stiffer shaft and a more flexible tip, enabling maneuverability with a monolithic design. To achieve this, a PC preform measuring 150 mm in length was designed with an inner diameter of 8 mm, an outer diameter of 24 mm, and equidistant side channels, each 4 mm in diameter (**Fig. 5**). The preform featured 10 longitudinally and periodically arranged segments, each 10 mm long, including a 2 mm section with a reduced outer diameter of 23 mm (**Fig. 5a and 5b**). The segmentation was designed to produce fibers with controlled stiffness and diameter, with thicker sections providing strength and thinner sections enhancing steerability.

During the thermal drawing process, the preform feed speed and fiber draw speed were maintained consistent. The transitions in diameter between the preform's segments played a critical role in achieving the intended mechanical properties. Specifically, the reduction from a 24 mm diameter to 23 mm an 8-9% decrease in radius generated localized stress variations. These stresses, amplified by the drawing process, ensured precise scale-down to the desired final dimensions, producing a thicker shaft of 2.2 mm and a more flexible tip of 1 mm. The cross-sectional features of the preform were preserved across different scale-down ratios, validating the robustness of the thermal drawing method.

After drawing, catheters were cut into lengths of 35 cm (**Fig. 5c**) and 120 cm (**Fig. 5d**). Four equidistant side channels allow wires to be fed through the holes and to manipulate the tip's motion by pulling these wires (**Fig. 5e**). Our approach simplified the fabrication process by producing several tip-only steerable catheters in one run, without extra post processing steps, demonstrating its efficiency for medical applications.

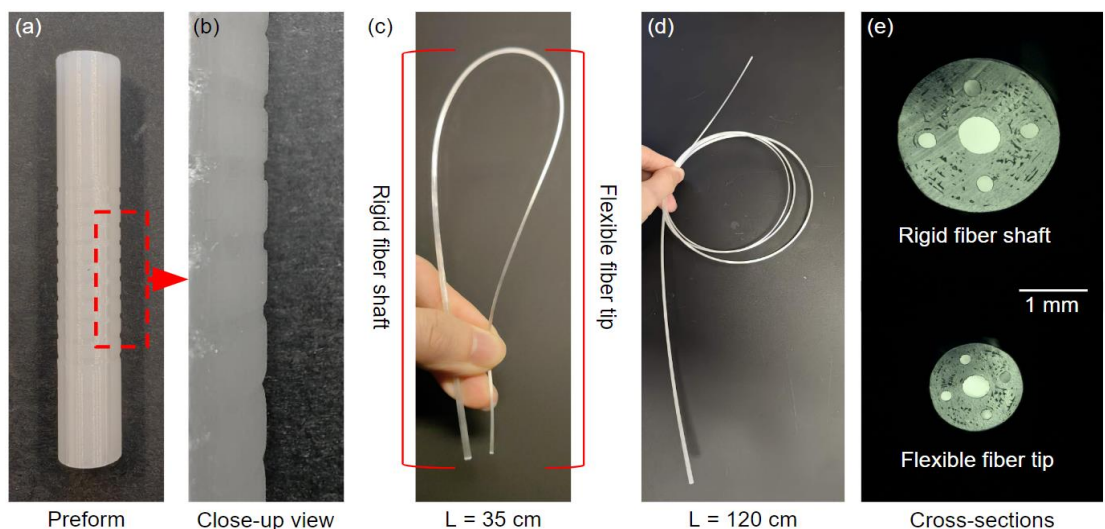

**Fig. 5 Preforms with longitudinal periodicity for tapered fibers.** (a) The preform features periodic segments along its length, designed to create fibers with variable diameters, and (b) a close-up view of the segments. (c) The fiber exhibits a rigid proximal shaft and a flexible distal end. (d) A longer version of the fiber demonstrates extended flexibility. (e) Cross-sectional views show equidistant holes for co-feeding wires, which enable manipulation of the fiber's flexible tip.

### 3.2.3 Fibers with Different Stiffness

Soft fibers have been thermally drawn for several applications, such as robotics and sensing. Most of fibers of these performs are fabricated by molding, which limits the complexity of design [29,31]. Expanding the material repertoire to include 3D-printed PP preforms, we successfully produced monofilament fibers through thermal drawing. According to the manufacturer's datasheet, the tensile stress values are 2.134 GPa for PC and 12 MPa for PP. Our approach, detailed in the Experimental section, utilized 3D-printed PP preforms to create flexible fibers without requiring post-processing steps such as laser profiling [35,51].

The mechanical properties of these fibers were evaluated by pulling molybdenum wires inserted through the side lumens of the fibers, with the wires fixed at the distal end. Tests were conducted on fibers of varying lengths, including 4, 6, and 8 cm for PP (**Fig. 6c**) and 8 cm for PC (**Fig. 6a**). The fiber bending angle as a function of the applied force for different fiber lengths was characterized (**Fig. 6b and 6d**).

PP fibers exhibited superior bending capabilities due to their lower tensile stress, presenting 5.1 times greater bending angles than PC fibers when same amount force applied (**Fig. 6e**). For 8 cm-long fibers, PP fibers conformed more closely to a circular shape with a tighter radius of curvature when compared to PC fibers. The 6 cm PC fiber kinked during the pulling test. In contrast, the flexibility of PP allowed tendons to bend a 4 cm fiber to a radius of curvature of 1.43 cm, highlighting their potential for applications requiring tight bending radius (**Fig. 6f**). This demonstrates the advantage of 3D printing in expanding the material repertoire for fiber fabrication, especially with challenging materials like semi-crystalline PP. Consequently, PP fibers can be suitable for instruments demanding higher flexibility, such as minimally invasive surgery devices.

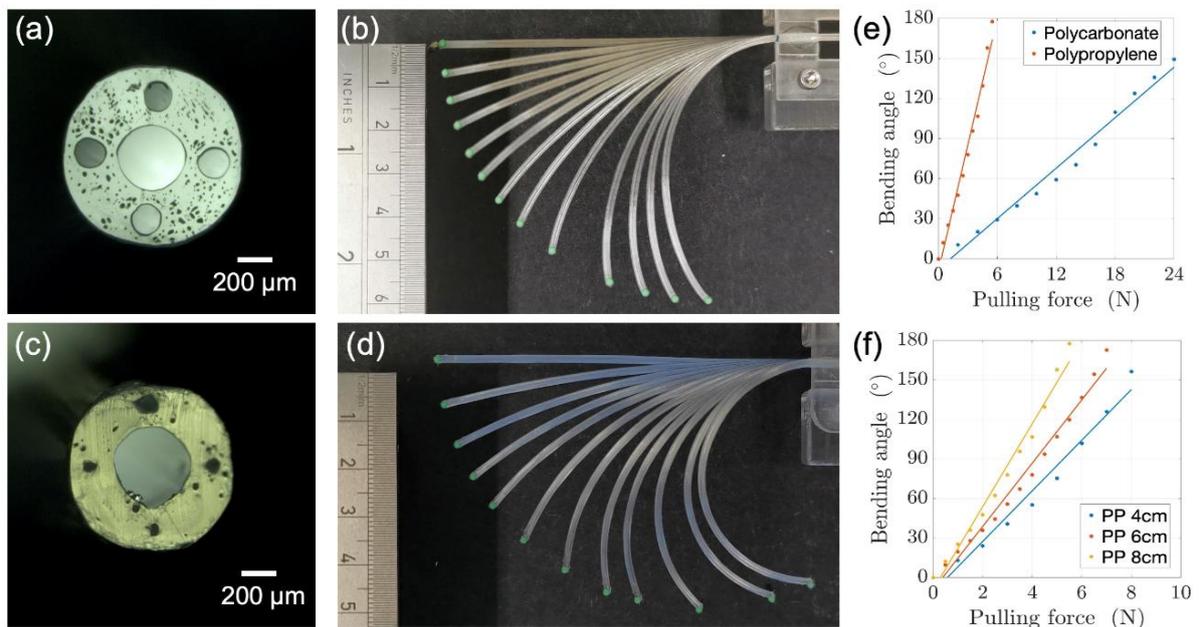

**Fig. 6 Mechanical testing of fibers for bending response.** (a) A PC fiber cross-section. (b) The bending response of an 8 cm long PC fiber with increasing applied force. (c) A polypropylene (PP) fiber cross-section. (d) The bending response of an 8 cm long PP fiber subjected to varying forces. (e) Comparison of the bending angles of PC and PP fibers relative to the pulling force. (f) Bending angles of PP fibers at different lengths to further show their flexibility at varying lengths (i.e., 4 cm, 6 cm, and 8cm).

### 3.2.4 Iron-Filled Multi-material Fiber for Magnetic Actuation

Magnetically actuated fibers are widely used in various fields, including soft robotics, medical applications, and flexible electronics. These fibers, often employed as flexible endoscopic tools for diagnostic and therapeutic purposes in complex surgical settings such as endovascular navigation, facilitate controlled movements in wearable devices and integrate smoothly into electronic fabrics [52–57]. Expanding on these developments, we designed and fabricated multi-material fibers combining PC and ferromagnetic polylactic acid (mPLA), as shown in **Fig. 7**. The first design uses a concentric approach with a central working channel suitable for passing medical tools (**Fig. 7a-c**). The second design integrates mPLA segments radially around three equidistant channels, enhancing magnetic interaction capabilities (**Fig. 7d** and **7e**). Details about materials and settings can be found in the experimental section.

Initially, we aimed to fabricate a fiber with concentric layers, where mPLA is encased within a PC shell (**Fig. 7a**). The isometric view of the preform confirms the precise concentricity (**Fig. 7b**). We successfully drew this preform into fibers of varying diameters, maintaining the structural fidelity of the internal channels and mPLA segments (**Fig. 7c**). Transitioning to more complex configurations, we designed a fiber with multiple mPLA sections embedded within the PC (**Fig. 7d**). The cross-section of the resulting fiber is shown in **Fig. 7e**.

To validate the magnetic properties, we tested the fibers by attracting them with a permanent magnet. The setup consisted of a fiber (100 mm in length and 1 mm in diameter) attracted by a permanent magnet (N35 Neodymium block, dimensions 40x25x30 mm) attached to a linear manual stage. This setup allowed us to move the permanent magnet closer to a magnetic fiber precisely. Sequential images captured the fiber's tip being attracted to the magnet, demonstrating noticeable deflection when the permanent magnet was approximately 20 mm away (**Fig. 7f**).

These prototype magnetic fibers validate the feasibility of employing thermal drawing techniques to produce magnetic fibers from 3D-printed preforms. The design of the fibers' cross-section and dimensions can be tailored to suit specific application requirements. For example, increasing the size of the mPLA component enhances the magnetic field generated by the fibers. The number and dimensions of the internal working channels can be customized to fit the specific tools intended for delivery.

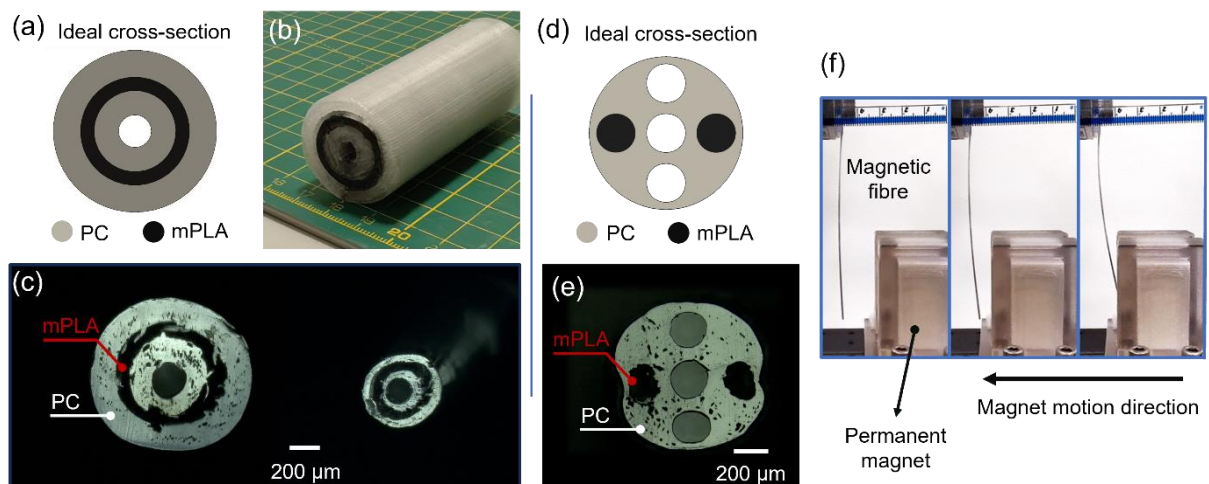

**Fig. 7 Fabrication and actuation of a magnetic multi-material fiber.** (a) The preform with an ideal fiber cross-section (b) featuring a central mPLA core within a PC sheath. (c) Cross-sections of the drawn fiber at varying diameters illustrating the consistency of the mPLA and PC configuration. (d) Another ideal cross-section design, this time with several mPLA sections embedded in a PC. (e) The corresponding drawn cross-section of the fiber where shrinkage near the mPLA areas is indicated by arrows, suggesting a reaction of the material during the drawing process. (f) Sequential images demonstrating the magnetic fiber's tip moving towards a permanent magnet validating the magnetic response and directional control of the fiber.

### 3.2.5 Hybrid Preform Fabrication Technique

The 3D printed preform approach, when combined with other preform fabrication methods can advance the properties or functions of the resulting fibers. As an example, we focused on Cyclo-Olefin Polymer (COP) and Cyclo-Olefin Copolymer (COC) polymers, which have good optical transparency in the visible and near-infrared (NIR) range light. These materials have been thermally drawn to fabricate polymeric optical fibers (POFs) [58–60]. However, both materials have been known to exhibit poor mechanical properties considering their brittleness [61,62]. Here, we demonstrated the feasibility of the hybrid preform approach to enhance mechanical properties of COC and COP fibers by adding the addition of an outer cladding.

We employed COP and COC bulk material tubes with an 8 mm inner diameter and a 20 mm outer diameter as the core, and 3D printed PC preforms with a 20 mm inner diameter and a 30 mm outer diameter as the outer shell. We assembled the bulk tube inside the 3D printed shell to form the fiber preform. This preform was subsequentially consolidated in a vacuum oven at 200 °C for 30 minutes and then co-drawn to produce multi-material fibers. Micro-CT imaging demonstrated that the cross-sectional integrity was maintained throughout, with no gaps between the materials (**Fig. 8a**). Figure 8b shows the cross-section quality of the COP and PC combination. Additionally, we drew COC and black PC combinations, as seen in the isometric (**Fig. 8c**) and top-view (**Fig. 8d**) microscope images. These results demonstrate the seamless integration of two materials in a single fiber drawn from a hybrid preform. This technique allows for efficient and cost-effective production of optical fibers with improved mechanical properties.

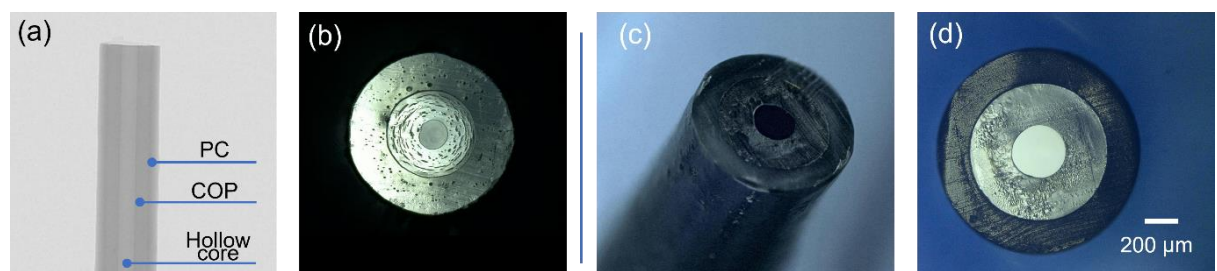

**Fig. 8 Visualization of fiber fabrication using 3D printed and bulk materials through micro-CT imaging and microscopy.** (a) A front-view micro-CT image and (b) A top-view image illustrates the preform comprising PC and COP. (c) An isometric and (d) a top-view microscope image of a fiber made with PC and COC.

## 4. Conclusions

This study explored the fabrication of advanced fibers with complex geometries, material compositions, and tailored functionalities by the introduction of additive manufacturing into the thermal drawing process. By utilizing 3D printing for preform fabrication, we were able to rapidly design and produce low-cost preforms with complex cross-sections and various materials. These preforms were thermally drawn to fabricate fibers with diameters as small as 200 µm, while preserving their cross-sectional features.

We characterized the effects of fabrication parameters on fiber quality and used the optimal settings to produce various preform and fiber designs. The preform designed with a jigsaw puzzle piece shape enabled the fabrication of tip-steerable fibers for neurosurgical applications. This example demonstrates how the proposed approach can easily overcome the challenging tasks of fabricating highly complex miniature structures. The preform with periodically varying diameters allowed for the production of multiple tapered fibers in a single run, which could be utilized for the scalable production of steerable catheters. Additionally, we applied this approach to fabricate fibers from various materials. We 3D-printed PP preforms and, for the first time, demonstrated the thermal drawing of this material into fibers. The low stiffness of PP is well-suited for steerable catheters requiring a low radius of curvature. We also demonstrated that, in addition to the polymers, we can introduce materials that response to external stimuli. The integration of mPLA into the 3D-printed preforms enabled the fabrication of fibers that can response to magnetic field, offering a novel approach for manufacturing magnetically actuated fibers.

The combination of 3D printing and thermal drawing technology enables rapid prototyping of fibers with bespoke designs, complex cross-sectional geometries, diverse material selections, and diameters scaled down to a few hundred micrometers. This low-cost fabrication approach overcomes several fabrication challenges, accelerating and advancing fiber research that can support innovation in various fields, such as healthcare and robotics. Future studies could explore additional material combinations, fiber architectures, and post-drawing functionalization techniques to further expand the scope and potential of this technology.

**Conflicts of interest:**
The authors declare no conflict of interest.

**Acknowledgments:**
The authors would like to thank Libaihe Tian for her support with photos and illustrations.

**Credit authorship contribution statement**
**Demircali Ali Anil:** Conceptualization, Data curation, Formal Analysis, Investigation, Methodology, Software, Validation, Visualization, Writing – original draft. **Zhao Jinshi:** Data curation, Formal Analysis, Methodology, Software, Validation, Visualization, Writing – original draft. **Aktas Ayhan:** Data curation, Visualization, Writing – original draft. **Abdelaziz Mohamed EMK:** Methodology. **Temelkuran Burak:** Funding acquisition, Project administration, Resources, Writing – review & editing